# Multimodal Transformer Models for Turn-taking Prediction: Effects on Conversational Dynamics of Human-Agent Interaction during Cooperative Gameplay

Young-Ho Bae[a], Casey C. Bennett[b,*]

[a]Department of Data Science, Hanyang University, Seoul, Korea
[b]Department of Computing & Digital Media, DePaul University, Chicago, IL, USA

**Abstract**

This study investigates multimodal turn-taking prediction within human-agent interactions (HAI), particularly focusing on cooperative gaming environments. It comprises both model development and subsequent user study, aiming to refine our understanding and improve conversational dynamics in spoken dialogue systems (SDSs). For the modeling phase, we introduce a novel transformer-based deep learning (DL) model that simultaneously integrates multiple modalities - text, vision, audio, and contextual in-game data to predict turn-taking events in real-time. Our model employs a Crossmodal Transformer architecture to effectively fuse information from these diverse modalities, enabling more comprehensive turn-taking predictions. The model demonstrates superior performance compared to baseline models, achieving 87.3% accuracy and 83.0% macro F1 score. A human user study was then conducted to empirically evaluate the turn-taking DL model in an interactive scenario with a virtual avatar while playing the game *Don't Starve Together*, comparing a control condition without turn-taking prediction (n=20) to an experimental condition with our model deployed (n=40). Both conditions included a mix of English and Korean speakers, since turn-taking cues are known to vary by culture. We then analyzed the interaction quality, examining aspects such as utterance counts, interruption frequency, and participant perceptions of the avatar. Results from the user study suggest that our multi-modal turn-taking model not only enhances the fluidity and naturalness of human-agent conversations, but also maintains a balanced conversational dynamic without significantly altering dialogue frequency. The study provides in-depth insights into the influence of turn-taking abilities on user perceptions and interaction quality, underscoring the potential for more contextually adaptive and responsive conversational agents. This research contributes significantly to the fields of HAI and SDSs by presenting a robust model for turn-taking prediction and shedding light on the practical applications and user perceptions of these systems.

[*]Corresponding Author
 *Email address:* cabennet.iu@gmail.com (Casey C. Bennett)

## 1. Introduction

Effective turn-taking is a fundamental component of human dialogue, facilitating smooth communication between two or more parties by preventing overlaps, interruptions, and inappropriate gaps that can disrupt the flow of information [1]. While humans typically manage this process effortlessly in everyday conversations, conversational systems (e.g., voice assistants, social robots) often face challenges during such natural conversations, including frequent interruptions and delayed responses, which can lead humans to feel like conversations with those artificial agents are less-immersive than conversations with other humans [2]. Despite extensive research, developing a comprehensive, continuous, and context-sensitive approach to address the diverse aspects of turn-taking for artificial agents remains a significant challenge [3, 4]. Such aspects include predicting whether a speaker will continue after brief silences, identifying moments for providing backchannels [5], estimating speaker response durations, and interpreting pauses in system-generated speech [6]. Spoken dialogue systems (SDSs), which are purposely designed for human-like interactions with artificial agents, require models to deal with those challenges in order to accurately determine appropriate moments for the system to initiate speech. Such artificial agents may include robots, virtual avatars, and chatbots, all of which fall under the broader fields of human-agent interaction (HAI) and human-robot interaction (HRI). However, there is still much room for improvement in those turn-taking models for SDSs.

In this paper, we propose a continuous and efficient approach for multi-modal turn-taking prediction using a transformer deep learning (DL) model. Our approach is specifically tailored for dynamic and context-rich human-agent dialogues, particularly in cooperative game environments where two parties must collaborate to achieve some tasks. The model predicts an agent's upcoming speech opportunity within a near-future time window (e.g., next few seconds), utilizing data from the preceding interaction period. We employ a sliding window technique for incremental data processing, simultaneously integrating multiple modalities: text, vision, audio, and contextual in-game information. These modalities are essential for comprehending and predicting turn-taking behavior during cooperative tasks between humans and artificial agents. Moreover, the computational efficiency of this approach is vital for real-time applications such as chatbots, gaming, virtual assistants, and real-world HRI in user homes and workspaces, where computational resources and response times are critical. To put it more simply, the system must respond quickly (within a second) if we intend it to mimic natural human conversation, so there is a trade-off between prediction accuracy and speed/computational complexity here.

To empirically assess the impact of such advanced turn-taking ability onboard an artificial agent, as facilitated by our proposed DL model, we established two experimental conditions: Turn-taking and Control. Only the former condition incorporated our model, whereas the latter (Control) condition did not. Both conditions were conducted within the same cooperative game environment, where a human and the artificial agent had to collaborate to "win" the game, during which we collected data about the spoken dialogue. We then employed three distinct analysis methods on the collected data, including analyzing utterance counts, interruption frequency, and survey instrument data, to evaluate the effectiveness of our turn-taking approach across different languages (Korean and English). The experiment was conducted in those two different languages, since turn-taking cues are known to vary across cultures (see Section 2.5).

In the following sections, we first review related work on turn-taking in dialogue in general as well as provide a detailed description of our turn-taking modeling approach (multi-modal transformer-based). That description includes details of various components (e.g. feature extraction for each input modality) in the model and the overall architecture of the proposed model. We then detail an initial benchmarking study and subsequent *in vivo* user study with human participants, discussing the implications of the results regarding turn-taking models for SDS and artificial agents more broadly. This work is intended to contribute to the development of more natural, efficient, and adaptable turn-taking prediction models for SDSs and conversational interfaces, with potential applications across various domains.

## 2. Related Work

Turn-taking prediction has been a significant area of research within the field of SDSs, with numerous studies attempting to address its challenges. In this section, we review some of the most pertinent studies on turn-taking prediction, highlighting the findings and limitations identified in these studies, which have inspired our own research here.

### 2.1. Conversational Factors in Turn-taking Research

Accurately modeling the various factors influencing conversational dynamics is a key challenge in turn-taking prediction. For instance, some research has focused on detecting instances when users might wish to resume speaking after brief pauses, a crucial factor in preventing unintended interruptions [7, 8]. Another significant factor of turn-taking involves identifying appropriate moments for providing brief feedback, known as backchannels, during ongoing speech [5]. Examples of backchanneling includes expressions like "mhmm" or "I see" that let the speaker know the other party is listening and/or understanding what is being conveyed.

Additionally, studies have explored the distinction between longer utterances and short responses from human listeners during the act of backchanneling, with longer utterances typically serving a different backchanneling purpose (e.g., expressing enthusiasm) but also leading to a greater degree of speech overlaps and/or interruptions. This differentiation can significantly impact how we think about designing the timing of artificial speech system responses [9]. In other words, some interruptions may be desirable at certain moments (e.g. wanting to convey enthusiasm). Furthermore, the interpretation by human users of pauses in an artificial system's speech has been thoroughly examined, given that pauses may have different contextual meanings at different times. For instance, researchers have investigated whether such pauses are perceived as turn-yielding or non-turn-yielding in different scenarios, underscoring the influence of subtle cues on speech synthesis during conversational flow [10, 11]. While these studies have primarily addressed individual factors of turn-taking, they collectively highlight the necessity for a unified and context-sensitive approach. Such an approach would more accurately capture the complexities inherent in conversational interactions, addressing the multifaceted nature of turn-taking dynamics.

*2.2. Limitations of ASR-Dependent Dialogue Systems*

Dialogue systems commonly rely on automatic speech recognition (ASR) systems, often integrated with voice activity detection (VAD), to determine the end of speech segments. These systems typically use preset thresholds to identify when a speaker has finished speaking. However, this reliance on ASR and fixed thresholds introduces significant challenges in effectively managing turn-taking dynamics. A primary concern with preset thresholds in ASR systems is achieving an optimal balance between responsiveness and accuracy. A threshold that is too short may cause the system to interrupt the user prematurely during natural speech pauses, thereby disrupting the conversational flow. Conversely, a threshold that is too long can make the system appear unresponsive or slow. This may result in the system delaying the recognition of the end of the user's turn, leading to unnatural pauses that detract from the fluidity of the dialogue [12].

Additionally, the dependency of dialogue systems on the accuracy of ASR presents further challenges. ASR systems frequently struggle to accurately detect speech in noisy environments during fast-paced conversational contexts, as well when humans use accents and/or dialect rather than "standard" language [13]. Misinterpretations or missed cues by the ASR can lead to incorrect turn-taking decisions, negatively affecting the overall user experience [14].

*2.3. IPU-based Approach and Continuous Turn-taking Models*

Inter-pausal unit (IPU) based models are widely adopted for turn-taking predictions in SDSs [15, 16, 17]. These models segment speech into continuous segments, where silences between speech (identified using VAD) do not exceed a predetermined duration threshold. The primary objective of IPU-based models is to determine whether a speaker will continue speaking or yield their turn at the end of an IPU. However, this approach assumes that turn-taking exclusively occurs when speech ceases, overlooking the common occurrence of overlapping speech in natural dialogues [4].

In contrast, continuous turn-taking models offer a dynamic alternative to IPU-based models. These models employ incremental processing of dialogue data [18, 19, 20], handling user input in small segments, such as frame-by-frame or word-by-word. This approach enables the system to make continuous turn-taking decisions. Continuous models excel at predicting when turns are likely to be completed, identifying optimal moments for overlapping backchannels, and deciding when to interrupt the user, thereby enhancing the natural flow of dialogue [2]. Moreover, incremental processing effectively addresses the limitations of IPU-based models that rely on VAD. VAD systems, which use fixed thresholds to determine the end of speech segments, may not accurately capture the nuances of turn-taking dynamics. In contrast, continuous models, as exemplified by the work of Skantze [21], process audio on a frame-by-frame basis and predict future speech activity, showing improved performance over traditional models and even human judges.

Overall, continuous models represent a more sophisticated approach to managing turn-taking in human-agent dialogues, averse to simply trying to identify pauses as in IPU-based models. They provide more responsive and adaptable interactions, aligning more closely with the complexities of natural human communication.

## 2.4. Incorporation and Fusion of Multimodal Information

Another significant challenge in turn-taking prediction is the incorporation and effective fusion of multimodal information, such as both verbal and non-verbal cues. Given that human communication is inherently multimodal, integrating both auditory and visual cues has been shown to improve turn-taking prediction performance. Morency et al. (2011) proposed a multimodal approach combining auditory and visual information, yielding enhanced performance compared to single-modality approaches that focus on only one or the other [22]. Similarly, Al Moubayed et al. (2008) emphasized the critical role of multimodal cues in turn-taking prediction during HRI and HAI, noting a marked improvement in accuracy when information from various modalities is utilized [23].

The effective fusion of these distinct modalities is essential, as each modality offers unique social cues and, when combined, can complement each other to offer a more comprehensive understanding of conversational dynamics. The challenge lies in doing so in a computationally efficient manner, of course. From a machine learning prospective, Baltrušaitis et al. (2018) offer valuable insights for integrating these heterogeneous data sources, thereby maximizing the strengths of each modality [24]. Furthermore, Atrey et al.'s (2010) research on multimodal fusion for next-generation video applications demonstrates the potential of combining diverse data streams to augment system performance across a variety of tasks [25].

Nevertheless, the process of fusing multimodal data is still fraught with many challenges. It entails not only the efficient integration of these modalities (for real-time processing) but also the adept handling of asynchronous and missing data, as discussed by Wang and Dash (2007) in their work on fusion strategies for multimedia information retrieval [26]. Additionally, the synchronization and alignment of audio-visual data coming from separate sensors, as investigated by Noulas and Englebienne (2011), further illustrate the intricacies involved in multimodal fusion, particularly in real-time applications [27].

*2.5. Language Differences in Turn-taking Cues*

Several studies have highlighted key differences in turn-taking practices across various languages, with these variations often reflecting underlying cultural norms. For instance, Zhu and Boxer's (2021) study comparing American English and Mandarin Chinese revealed distinct approaches to managing prolonged simultaneous speaking and strong disagreements (i.e. conflicts of opinion), reflecting the cultural norms of each language [28]. That study underscored the diversity in turn-taking strategies and perceptions of politeness across different linguistic contexts. Beyond that, Levinson and Torreira's (2015) work explored the timing aspects of turn-taking and its implications for language processing [29]. Their work challenged existing language processing theories, suggesting that turn-taking patterns adhere to specific rhythms that are *idiosyncratic* to each language. This necessitates a more nuanced understanding of the complex dynamics in conversational timing. However, there are some commonalities across languages. For instance, Stivers et al. (2009) conducted a comprehensive study on turn-taking in conversations across a variety of cultures, covering 10 languages worldwide [30]. Their research identified a universal tendency to avoid overlapping speech and to minimize pauses between turns. This finding suggests the existence of common turn-taking practices across different cultural backgrounds, indicating certain universal aspects in conversational dynamics.

## 3. Methods

*3.1. Virtual Avatar & Speech System*

We previously developed an autonomous 'Social AI', a virtual avatar integrated with a speech system that was designed to facilitate HAI in a cooperative survival game, as detailed in Section 3.2. The goal was to evaluate various factors affecting speech interactions between humans and artificial agents. The development of this Social AI was initially informed by insights from recordings of human vs. human gameplay. That was followed by subsequent evaluations of the various developed speech components on an artificial agent, which have significantly improved its interactive speech capabilities over time, as elaborated in our previous works [31, 32, 33].

The Social AI features a diverse range of speech utterances, categorized into 46 distinct groups. Each category corresponds to different in-game situations, such as resource collection, combat, or strategy planning, and is organized in a multi-level hierarchy structure. The utterances include both self-generated utterances, based on the Social AI's internal logic, as well as responses to the human player's speech, enabled by the ASR component. Moreover, the system is multilingual, capable of speaking in English, Korean and Japanese.

The speech system of the Social AI was custom-built using Python programming. It incorporates locally-installed voice packages for the Text-to-Speech (TTS) module, and channels audio outputs to an internal 'virtual' microphone jack. The ASR component, integral to the system's interactive capability, is powered by the Microsoft Azure Speech-to-Text (STT) API for human speech recognition in both English and Korean. The speech outputs are transmitted through the internal microphone jack to the Loomie application (https://www.loomai.com/), which visualizes the Social AI. This application enabled the creation of an ethnically neutral female avatar, a crucial element for our cross-cultural HAI study. The avatar's lip movements are synchronized with the speech output, enhancing the realism of the interaction.

*3.2. Cooperative Game Environment*

In the current study, we selected the video game *Don't Starve Together* as the cooperative game environment (https://www.klei.com/games/dont-starve-together), which is accessible via the Steam platform. This survival game challenges players to gather resources, craft tools, combat creatures, and collaborate to extend their survival. Similar to other social survival games, such as Minecraft, *Don't Starve Together* requires players to combine specific resources to create essential items, crucial for survival against increasing in-game threats. The game's structure, supporting both free-form play and goal-oriented tasks within its cooperative multiplayer modes, made it an ideal setting for our study. It encourages collaborative problem-solving and strategy development, aligning well with our research objectives.

*Don't Starve Together* is notable for its customizability, offering game modification tools that enable users to alter in-game environments and NPC behaviors using LUA programming. We used these tools to develop a specific

"Game Mod" for two primary purposes. Firstly, it was used to capture real-time game data, including player status, inventory, movements, time of day, and proximity to monsters or structures. Secondly, it enabled us to modify game settings to influence interactions between the avatar and human player. We configured the game to start in a resource-rich area ("advanced start"), provided a permanent light source at the starting position to encourage players to return to it periodically ("base camp"), and adjusted settings to reduce player deaths, aiming for each session to last approximately 30 minutes ("partial invincibility"). These game modifications were not disclosed to the participants.

The Game Mod functioned concurrently with the avatar's speech system (i.e., Social AI), utilizing the real-time game data to contextualize the avatar's speech. Furthermore, the collected game data was primarily utilized as one of the modalities for training our proposed turn-taking transformer DL model, as detailed in Section 3.3.2.

### 3.3. Model Development & Benchmarking Study
### 3.3.1. Turn-taking Model Design

Prior to conducting the user study (see Section 3.4 below), we first developed a multimodal transformer DL model tailored for enhancing turn-taking in HAI within a cooperative game environment, that was trained/tested in a benchmarking study of previously collected HRI experimental data. The model predicts when a virtual avatar should initiate speech by analyzing data from the previous five seconds to forecast turn-taking events in the upcoming second. This approach takes into account both the latency of language production, which typically ranges from 600 to 1500ms, and the much shorter typical gaps between turns in conversation, which are often as brief as 100 to 300ms [20, 29]. By predicting turn-taking events one second in advance, our model aims to reconcile these two distinct timescales, creating a more natural interaction rhythm. The real-time prediction is made possible through a sliding window technique, processing a blend of text (T), vision (V), audio (A), and in-game (G) context data.

### 3.3.2. Feature Extraction

Data sources include OBS recordings (audio and vision), conversation transcripts (text), and custom game modifications (in-game context). To optimize computational efficiency, we employ simplified, pre-trained DL models for text and vision data processing, the openSMILE toolkit for audio feature extraction, and varying sample rates aligned with each data type's characteristics. This approach ensures swift processing and CPU conservation, making the system suitable for real-time applications in resource-limited environments. Each modality undergoes specialized feature extraction, with the resulting features integrated into a higher-level representation for the predictive model. The model outputs a probability indicating the avatar's likelihood of appropriately initiating speech in the following second.

***Text features***. Verbal cues are crucial for turn-taking in conversational interactions [3, 9, 30]. We sampled dialogues from both the human player and the avatar within five-second windows during gameplay (see Figure 1). For text processing, we employed the cased multilingual DistilBERT (DistilmBERT) model [34], suitable for bilingual (English and Korean) speech processing. DistilmBERT, with 134 million parameters and 6 transformer layers, offers comparable performance to mBERT while being 24% smaller and twice as fast. The process involves concatenating the utterances from both the avatar and human player along with their timestamps, tokenizing the combined text using WordPiece [35], and then transforming it into 768-dimensional contextualized vector representations:

$$t_i = \{[CLS], \omega_1, \omega_2, \ldots, \omega_n, [SEP]\}$$
$$F_t = \text{DistilmBERT}(t_i, \theta_t) \in \mathbb{R}^{d_t}, \quad i \in [1, n] \quad (1)$$

where $t_i$ represents the tokenized text sequence, starting with [CLS], a special token for classification tasks, and ending with [SEP], a separator token. The sequence also includes subword tokens $\omega_1, \omega_2, \ldots, \omega_n$ from the concatenated dialogue. $F_t$ denotes the feature representation output by DistilmBERT for each tokenized input, where $\theta_t$ are the model's hyperparameters and $d_t$ represents the dimensionality of the output feature space, which in this case, is 768.

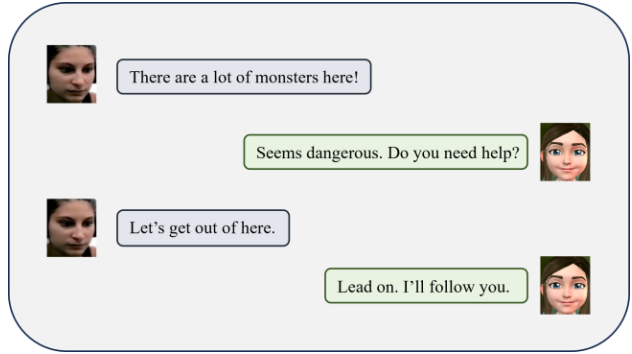

Figure 1: Dialogue Example during Experiment

***Vision features***. Visual cues such as facial expressions are fundamental to non-verbal communication [36, 37]. For vision features, we focused solely on capturing and extracting the human player's facial expressions. We utilized the DeepFace toolkit (https://github.com/serengil/deepface) with MobileFaceNets optimized by the Sigmoid-Constrained Hypersphere Loss (SFace) function [38]. These efficient CNN models, with fewer than 1 million parameters, balance size, speed, and accuracy for real-time applications. The SFace function ensures intra-class compactness and inter-class separability, producing 128-dimensional visual embeddings of facial expressions every second.

*Audio features*. We employed the openSMILE (https://www.audeering.com/) toolkit's ComParE 2016 feature set [39] for audio analysis. This set comprises 65 low-level descriptors (LLDs) categorized into spectral, energy, voicing-related, and cepstral features, as described in Table 1. Using a 50ms frame size, we extracted features and averaged them over half-second intervals, resulting in 10 sequences per 5-second audio segment. This approach balances granularity and computational efficiency for real-time applications. The extracted LLDs are then normalized to produce the final audio embedding vector.

*In-game features*. Our custom Game Mod for *Don't Starve Together* includes a 'game data logging' feature that captures the game state at one-second intervals in real-time. We selected 40 key data points that provide a comprehensive view of both the human player's and avatar's in-game status (see Table 2). These data points cover aspects such as health levels, inventory contents, movement patterns, combat engagements, and proximity to in-game entities, characterizing the dynamic game context.

Table 1: ComParE 2016 Feature Set

| 4 energy related LLD | Group |
|---|---|
| Sum of auditory spectrum (loudness) | prosodic |
| Sum of RASTA-filtered auditory spectrum RMS | prosodic |
| Energy, Zero-Crossing Rate | prosodic |
| **55 spectral LLD** | **Group** |
| RASTA-filt. aud. spect. bds. 1–26 (0–8 kHz) | spectral |
| MFCC 1–14 | cepstral |
| Spectral energy 250–650 Hz, 1 k–4 kHz | spectral |
| Spectral Roll-Off Pt. 0.25, 0.5, 0.75, 0.9 | spectral |
| Spectral Flux, Centroid, Entropy, Slope | spectral |
| Psychoacoustic Sharpness, Harmonicity | spectral |
| Spectral Variance, Skewness, Kurtosis | spectral |
| **6 voicing related LLD** | **Group** |
| F0 (SHS & Viterbi smoothing) | prosodic |
| Prob. of voicing | voice qual. |
| log. HNR, Jitter (local & $\delta$), Shimmer (local) | voice qual. |

*3.3.3. Proposed Model Architecture*

Figure 2 illustrates our proposed model architecture for turn-taking prediction. The model comprises four parallel processing pipelines, each corresponding to a different data modality: text, vision, audio, and in-game context. Each pipeline begins with a 1D convolutional layer and a bidirectional LSTM (Bi-LSTM) for initial encoding. These encoders transform the raw data into higher-level feature representations. A Crossmodal Transformer then refines these representations by integrating cues across different modalities. Following the transformer, an additional Bi-LSTM layer further processes its output. Finally, the outputs from all pipelines are concatenated to form a comprehensive feature set, which is then fed into a sigmoid layer for final turn-taking predictions.

Table 2: Game Data Description

| General Data | Description |
|---|---|
| Phase | Current phase in the game |
| Distance | Distance between Avatar and Player |

| Entity Data (Avatar, Player) | Description |
|---|---|
| Hunger | Current hunger level |
| Health | Current health level |
| Sanity | Current sanity level |
| Xloc | X-coordinate location in the game world |
| Zloc | Z-coordinate location in the game world |
| Curr Inv Cnt | Number of items in the current inventory |
| Curr Active Item | Current active item |
| Curr Equip Hands | Item equipped in the hands |
| Attack Target | Current attack target |
| Defense Target | Current defense target |
| Recent Attacked | Recently attacked status |
| Food | Current food status |
| Is Light | Light status |
| Monster Num | Number of monsters encountered |
| Twig | Number of twigs in the inventory |
| Flint | Number of flints in the inventory |
| Log | Number of logs in the inventory |
| Rock | Number of rocks in the inventory |
| Grass | Number of grasses in the inventory |

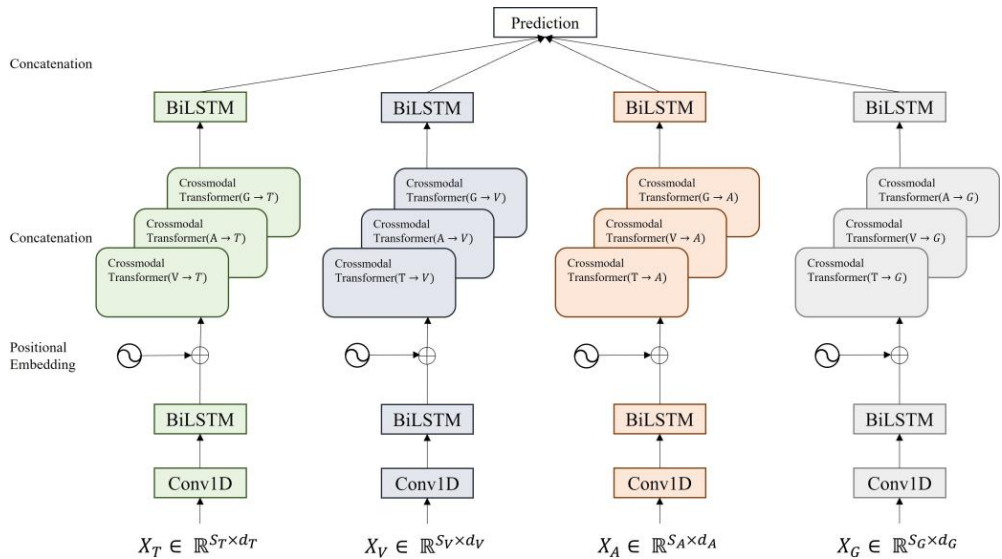

Figure 2: Overall Model Architecture

***Temporal Convolution***. Following feature extraction (Section 3.3.2), modality- specific embedding vectors $X_{(T,V,A,G)} \in \mathrm{R}^{S_{(T,V,A,G)} \times d_{(T,V,A,G)}}$ are processed through 1D temporal convolution layers. Here, $S$ represents the sequence lengths for each modality. These layers employ learnable filters to extract temporal patterns while preserving sequential order [40, 41, 42]. This process aligns features from different modalities into a common dimensional space $d$, facilitating integration in the subsequent Crossmodal Transformer layer:

$$D_{\{T,V,A,G\}} = \mathrm{Conv1D}(X_{\{T,V,A,G\}}, K_{\{T,V,A,G\}}) \in \mathrm{R}^{S_{\{T,V,A,G\}} \times d} \quad (2)$$

where $K_{\{T,V,A,G\}}$ represents the convolutional kernel sizes for text ($T$), vision ($V$), audio ($A$), and in-game ($G$) modalities.

***Bi-LSTM***. After 1D temporal convolution, $D_{\{T,V,A,G\}}$ are processed through modality-specific Bi-LSTM layers, capturing temporal dependencies by processing data bidirectionally, considering both past and future context [43, 44]:

$$B_{\{T,V,A,G\}} = \mathrm{BiLSTM}(D_{\{T,V,A,G\}}) \in \mathrm{R}^{S_{\{T,V,A,G\}} \times 2d} \quad (3)$$

where $B_{\{T,V,A,G\}}$ represents the Bi-LSTM hidden states for text ($T$), vision ($V$), audio ($A$), and in-game ($G$) modalities.

***Positional Embedding***. To provide temporal position information, we add sinusoidal position embeddings (PE) [45] to the Bi-LSTM outputs $B_{\{T,V,A,G\}}$:

$$C_{\{T,V,A,G\}} = B_{\{T,V,A,G\}} + \mathrm{PE}(S_{\{T,V,A,G\}}, 2d) \in \mathrm{R}^{S_{\{T,V,A,G\}} \times 2d} \quad (4)$$

This step ensures the transformer's attention mechanism can recognize input order, preparing the sequences for the Crossmodal Transformer.

***Crossmodal Transformer***. The core of our model comprises Crossmodal Transformers with $N$ layers of attention blocks (Figure 3). For example, transferring in-game ($G$) information to text ($T)$ is computed as:

$$\begin{aligned}
C_{G \to T}^{[0]} &= C_T^{[0]} \\
\hat{C}_{G \to T}^{[i]} &= MultiHead(LN(C_{G \to T}^{[i-1]}), LN(C_G^{[0]})) + LN(C_{G \to T}^{[i-1]}) \\
C_{G \to T}^{[i]} &= f_{\theta_{G \to T}^{[i]}}(LN(\hat{C}_{G \to T}^{[i]})) + LN(\hat{C}_{G \to T}^{[i]})
\end{aligned} \quad (5)$$

where $f_\theta$ is a feed-forward sublayer and LN is layer normalization [46]. This process, inspired by [47], models interactions between all modality pairs, resulting in 12 distinct Crossmodal Transformers for our four modalities (text, vision, audio, in-game). The outputs targeting the same modality are concatenated:

$$C_T = [C_{V \to T}; C_{A \to T}; C_{G \to T}] \in \mathbb{R}^{S_{\{T,V,A,G\}} \times 6d} \tag{6}$$

**Bi-LSTM and Prediction.** The concatenated outputs $C_{\{T,V,A,G\}}$ undergo further refinement through modality-specific Bi-LSTM layers:

$$L_{\{T,V,A,G\}} = \text{BiLSTM}(C_{\{T,V,A,G\}}) \in \mathbb{R}^{2d} \tag{7}$$

Here, $L_{\{T,V,A,G\}}$ denotes the final hidden state from the Bi-LSTM layers for text ($T$), vision ($V$), audio ($A$), and in-game ($G$) modalities, with a notable reduction in dimension to $2d$. These final outputs $L_{\{T,V,A,G\}}$ are concatenated into a unified feature vector, which is then processed through fully-connected layers with a sigmoid activation function to generate predictive probabilities. More details on the model architecture can be found in [48].

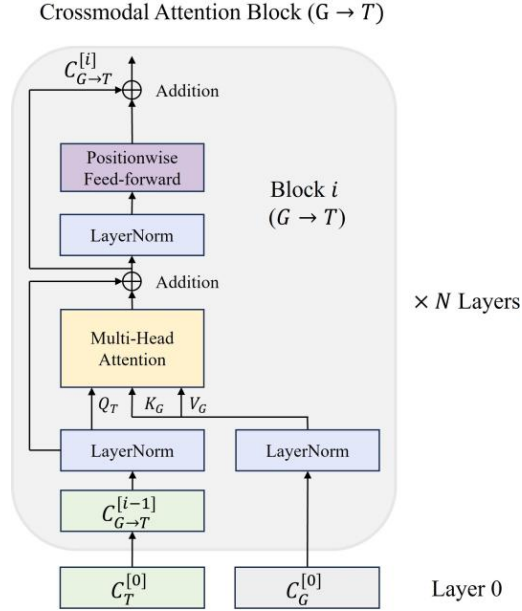

Figure 3: Crossmodal Transformer ($G \to T$)

*3.3.4. Turn-taking Event Determination*

Our research aimed to predict upcoming turn-taking events (i.e., "opportunities" to take a turn) within a binary classification framework, enabling the virtual avatar to identify optimal moments for initiating speech. The model's target variable is a binary label: 'class0' indicates a turn-taking event, while 'class1' denotes its absence.

To systematically determine turn-taking events, we initially computed speech segment durations and silence intervals (Inter-Pausal Units or IPUs) for both the avatar and human participant, thereby pinpointing potential turn-taking points [4]. Additionally, we incorporated speaker transitions and speech segment timing within a predetermined one-second window. This window was chosen to accommodate the typical latency of language production (600-1500ms) while allowing for the much shorter turn-taking gaps (100-300ms) observed in natural conversations [20, 29]. A turn-taking event is identified when a speaker transition occurs within this window, inspired by Skantze's method [21]. Using this method, the interruption situations can be effectively addressed, as they are naturally considered inappropriate interruptions.

Importantly, instances where neither the avatar nor the human participant speaks during the preceding 5 seconds are excluded, as they do not contribute to the model's understanding of turn-taking dynamics. Short speech segments detected by the ASR, even with speaker transitions, were classified as 'class1' (no turn-taking event) as these are considered conversational elements like backchannels.

It is crucial to note that while the model predicts turn-taking opportunities, the actual decision to take a turn relies on the Social AI's internal logic. The avatar refrains from speaking if the model indicates 'No turn-taking timing', even if ready to speak. Conversely, if the avatar is not ready to speak in a given situation, then obviously it doesn't speak regardless.

*3.3.5. Model Benchmarking Dataset*

The benchmarking dataset for training/testing the model comprised 126,886 samples from 60 human participants, derived from our previous experiments [31]. This dataset is distinct from that used in the Turn-taking and Control experiments (Section 3.4.2). While balanced between Korean and English data, it exhibited class imbalances of 82% and 71% respectively, with an overall 76% (96,433 samples) belonging to the majority class ('class1', indicating absence of turn-taking events). We allocated 80% for training and 20% for testing.

We opted to use human-agent dialogue data over human-human dialogue data, aligning with previous research [49, 50, 51, 52, 53]. This choice allows our model to learn from specific HAI successes and failures, including turn-taking events potentially overlooked by ASR systems. Moreover, HAI's unique turn-taking patterns differ from human-human interactions, enabling our approach to adapt and enhance user experience while considering computational limitations. By leveraging these interaction patterns, our model is better equipped to navigate conversational intricacies, advancing seamless human-machine communication.

*3.3.6. Model Training*

To address the imbalance problem in our dataset, we adopted the focal loss function [54] as our optimization criterion. The focal loss function is specifically designed to tackle class imbalance by down-weighting well-classified examples, thereby allowing the model to focus more on hard, misclassified examples. The focal loss is defined as follows:

$$FL(p_t) = -\alpha_t(1-p_t)^\gamma \log(p_t) \tag{8}$$

where $p_t$ is the true class probability, $\alpha$ is a balancing factor, and $\gamma$ is a focusing parameter.

We trained the model using mini-batches of 64 samples over 30 epochs, with an initial learning rate of 5e-5. The Adam optimizer was used for weight adjustment, and a sigmoid activation function was applied to the final layer. Table 3 details the hyperparameters used in our study.

Table 3: Hyperparameters of Our Proposed Model

| Hyperparameter | Value |
|---|---|
| Optimizer | Adam |
| Batch size | 64 |
| Initial Learning Rate | 5e-5 |
| Sequence Length ($T/V/A/G$) | 35/5/10/5 |
| Common space $d$ | 32 |
| Transformers Hidden Unit Size | 128 |
| Number of CM Blocks $N$ | 3 |
| Number of CM Attention Heads | 8 |
| 1D Conv Kernel Size ($T/V/A/G$) | 1/1/1/1 |
| Dropout | 0.1 |
| Bi-LSTM Hidden Unit Size | 64 |
| Number of Epochs | 30 |
| Focal loss $\gamma$ | 2 |
| Focal loss $\alpha$ | 0.3 |

*3.3.7. Model Optimization*

To optimize our model for resource-constrained environments, we employed TensorFlow Lite (TFLite) (https://github.com/tensorflow/tflite-

support), a framework designed for efficient on-device inference. TFLite operates through several key processes. We used the TFLite Converter to transform our TensorFlow model into a compact FlatBuffer format, aiming to reduce model size. Post-training quantization was applied, converting 32-bit floating-point weights to 16-bit floating-point representations, with the goal of reducing memory usage and computational requirements. TFLite automatically fused compatible operations to potentially reduce the number of computations and memory accesses during inference. Additionally, it utilizes hardware-specific optimizations, leveraging available CPU instructions to enhance execution speed.

The 16-bit quantization was specifically chosen to balance the trade-off between model size reduction and maintaining accuracy. This optimization approach was intended to improve CPU and hardware accelerator latency while preserving the model's performance. The specific outcomes of this optimization process, including its effects on model size, inference time, and performance metrics, are detailed in Section 4.1.3.

*3.3.8. Evaluation Metrics*

To comprehensively evaluate our proposed model on the dataset, we held out 20% of the benchmarking data for later testing, while the other 80% was used for model training. We then employed four key evaluation metrics: accuracy, macro F1-score, precision, and recall. Accuracy measures the overall proportion of correct predictions but can be misleading for imbalanced datasets. To address this limitation, we incorporated the macro F1-score, which provides an unweighted mean of F1-scores for each class, offering a more balanced assessment.

Precision and recall offer critical insights into the model's performance. Precision quantifies the proportion of true positive predictions among all positive predictions, while recall measures the model's ability to identify all relevant instances. This combination of metrics provides a nuanced and robust assessment of the model's performance, particularly in the context of an imbalanced classification dataset.

*3.4. User Study Experimental Design*

*3.4.1. Experimental Conditions*

To evaluate our turn-taking model beyond the validation testing during benchmarking described in Section 3.3 above, we conducted a follow-up user study with the turn-taking model deployed in real-time during conversations between an artificial agent and real humans. There were two distinct experimental conditions in the user study to assess the effects of our turn-taking model. Similar to the benchmarking study, the artificial agent took the form of an on-screen virtual avatar in the user study, which conversed with people within the same *Don't Starve Together* game environment.

The first condition, termed the 'Turn-taking' condition, was designed to investigate the effects of the Social AI's ability to predict appropriate moments for initiating speech, rather than speaking at arbitrary times, with real human users. This capability is derived from a multimodal DL model (detailed in Section 3.3) that processes turn-taking interaction cues, developed from data obtained in our previous experiments [31]. The 'Turn-taking' condition aimed to assess the model's practical application in real-time interactions. The second condition, referred to as the 'Control' condition, served as a baseline for comparative analysis. In that control scenario, the avatar's speech system operated using its default settings, without the integration of any advanced turn-taking prediction algorithm. This setup allowed us to observe the avatar's standard interactive behavior, thereby providing a direct comparison against which to measure whether interactions were enhanced by our proposed turn-taking model.

### 3.4.2. Participants

For the experiments conducted, we recruited 60 participants, who were divided into two groups corresponding to the experimental conditions: the Turn-taking group and the Control group. All participants were proficient in either English (at least TOEIC Level B2) or Korean (at least TOPIK Level C1 or "level 5"). The Turn-taking group comprised 40 participants, including 30 native Korean speakers residing in Korea and 10 English speakers. The English speakers were primarily university exchange students from North America and Europe (i.e., a mix of L1 and L2 speakers), currently studying in Korea. Originally the Turn-Taking condition was done all in Korean, with the English participants added later (via IRB addendum) to see if the results were the same or not across languages. The Control group consisted of 20 participants, evenly divided between 10 Korean and 10 English speakers. This group, similar to the Turn-taking group, included both native Koreans and foreign exchange students.

In both the Turn-taking and Control conditions, the participant demographics were approximately balanced in terms of gender (roughly 50/50), and the average age was 22.9 years. All ethical and procedural considerations, including the determination of sample size, were thoroughly reviewed and approved by the Hanyang University Institutional Review Board (IRB) under the code HYU-2022-281.

### 3.4.3. Experiment Setup & Procedure

For the experiment, we established two separate computer setups in different rooms. The 'player computer' was designated for the human participant, while the 'confederate computer' was utilized for controlling the virtual avatar. Both computers were connected to the same online game server to ensure synchronized gameplay. The player computer was equipped with an HD camera, headphones, and a Blue Snowball microphone, providing high-quality audio-visual input and output. Each participant engaged in a 30-minute game session with the virtual avatar on a private server, configured for 2-player cooperative mode. Prior to the session, a 5-minute tutorial was provided in the participant's native language to familiarize them with the game mechanics. For both the Turn-taking and Control conditions, the entire experimental session was conducted exclusively in one language, either Korean or English, depending on the participant. Additionally, a Zoom meeting was set up to facilitate direct audio and visual communication between the participant and the avatar during gameplay, with both displayed side-by-side on the screen, as illustrated in Figure 4.

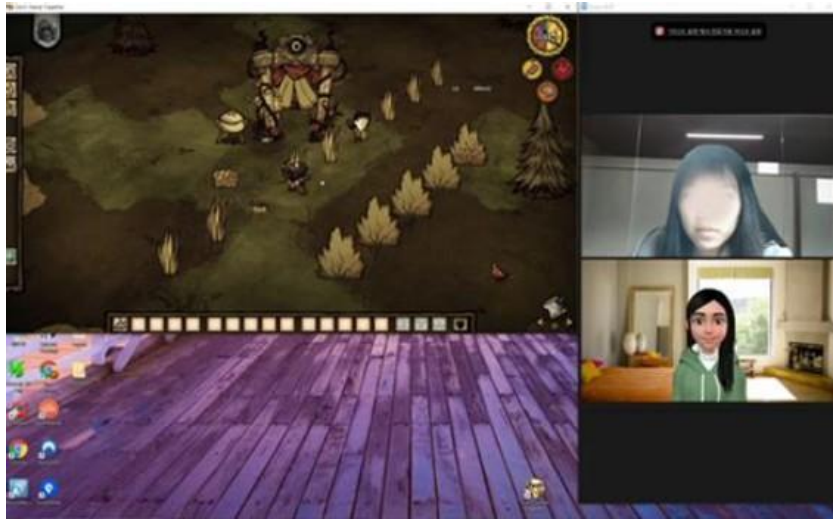

Figure 4: Gameplay example during experiment

### 3.4.4. Description of Collected Data

In each experiment, we systematically gathered three primary types of data: 1) audio-visual recordings of the gameplay, 2) detailed game data logs, and 3) instrument data on human perceptions. These data types were integral to comprehensively understanding the interactions between participants and the virtual avatar, ensuring consistency and reliability in our analysis across all sessions.

For the audio-visual recordings, we utilized OBS Studio (https://obsproject. com/) to capture the entire computer screen during gameplay. This included both the game window and the Zoom window, which displayed the players interacting in real time. This setup enabled us to synchronize the speech and reactions of both the avatar and the human player with the in-game events. We complemented these recordings with detailed game data, collected through the Game Mod, as detailed in Section 3.2, to facilitate an in-depth analysis of the interactive dynamics influenced by gameplay events.

Following each 30-minute game session, we evaluated the participants' perceptions and experiences using established HRI instruments. The Godspeed Questionnaire Series was used to assess general perceptions of the avatar, focusing on lifelikeness and perceived intelligence [55]. This instrument evaluates perceptions across five subscales: Anthropomorphism, Animacy, Likeability, Perceived Intelligence, and Perceived Safety. Each subscale consists of several items rated on a Likert scale, providing a comprehensive view of participants' attitudes and perceptions towards an artificial avatar. Its extensive use in HRI studies provides critical insights into the factors shaping HRI. Additionally, we employed the Networked Minds measure of social presence to gauge the immersive quality of the interactions [56, 57]. This instrument assesses the degree to which participants feel socially connected with an artificial avatar, focusing on aspects such as mutual awareness, attention allocation, and perceived emotional interdependence, also utilizing a Likert scale for responses. The comparative perspective provided by the Networked Minds instrument, with its roots in psychology, significantly enriches our analysis and deepens our understanding of the perceived realism and engagement during the sessions.

*3.4.5. Data Analysis Approach*

To analyze differences in language and experimental conditions, the initial step involved extracting speech data from the OBS recordings. We utilized Google Cloud Platform (GCP) services for speaker diarization, distinguishing between the speech of the avatar and the human participant, with timestamps allowing for synchronization with in-game events. A manual post-diarization cleanup was conducted to enhance the accuracy of the tran- scripts. Subsequently, we conducted a comprehensive analysis of the speech data, categorizing it by language and experimental condition. Specifically, we applied two-tailed independent-samples t-tests to assess differences across various experimental conditions within the same language experiments.

The analysis included calculating the utterance counts for both human participants and avatars. Additionally, an interruption analysis was performed, where we identified instances of either the avatar or the human participant speaking over the other. This process involved manual annotation of the speech transcript data, pinpointing moments where utterance timestamps overlapped without any intervening pause, indicating an interruption. Lastly, we analyzed instrument data to identify variations in human perceptions of the avatar during gameplay.

## 4. Results

*4.1. Model Performance - Benchmark Testing*

*4.1.1. Benchmarking Results*

We evaluated our proposed multimodal turn-taking prediction model against several baseline models: Early Fusion LSTM (EF-LSTM), Late Fusion LSTM (LF-LSTM), and Multimodal Transformer (Mult) [47]. All models were trained on identical text, vision, audio, and in-game data from our previous experiments with 60 human participants (i.e. our benchmarking dataset) [31]. To ensure a fair comparison, we maintained a roughly consistent number of parameters across all models. Each model underwent 10 training iterations with identical hyperparameter configurations.

Compared to the baseline models, our model demonstrated superior performance across all metrics on the benchmarking dataset (see Table 4), achieving the highest accuracy of 87.3% and macro F1 score of 83.0%. This represents a significant improvement over the best baseline model (Mult), which achieved 82.8% accuracy and 77.0% macro F1 score. These results underscore our model's potential to enhance turn-taking in HAI. Additionally, it exhibited faster convergence during training compared to the baselines (Figure 5), further highlighting its efficiency and effectiveness.

Table 4. Model Performance on Benchmarking Dataset

| Metric  | Acc  | Macro F1 | Precision | Recall |
|---------|------|----------|-----------|--------|
| EF-LSTM | 82.7 | 76.9     | 76.0      | 78.2   |
| LF-LSTM | 81.7 | 75.2     | 74.6      | 75.9   |
| Mult    | 82.8 | 77.0     | 76.2      | 78.1   |
| Ours    | 87.3 | 83.0     | 81.9      | 84.2   |

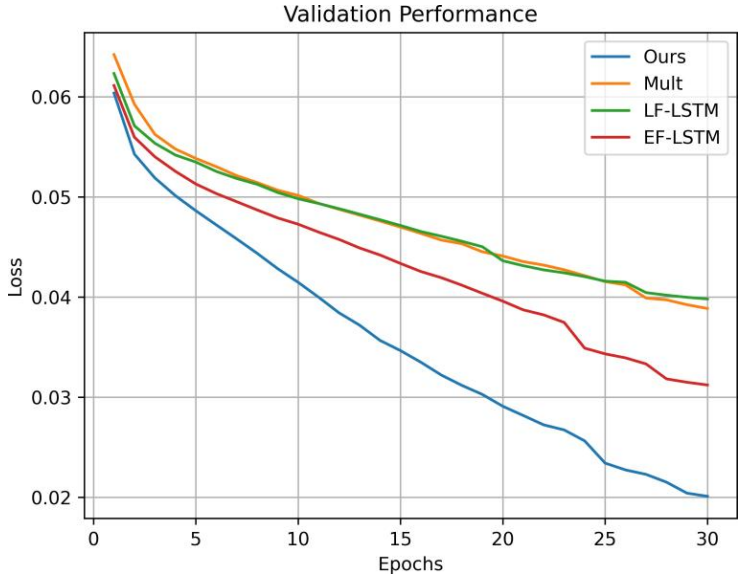

Figure 5: Validation Loss Across Training Epochs

*4.1.2. Ablation Study*

We conducted an ablation study to assess the contribution of individual components within our model, e.g. text, audio, vision, in-game data (see Table 5). Unimodal encoder evaluation revealed that text and audio modalities has the most significant impact on turn-taking prediction, outperforming vision and in-game contextual data. Specifically, the text-only model achieved the highest unimodal performance with 79.9% accuracy and 72.7% macro F1 score, followed closely by the audio-only model with 79.4% accuracy and 72.1% macro F1 score. In contrast, vision-only and in-game-only models performed notably lower, with accuracies of 73.4% and 73.3% respectively.

Examination of each Crossmodal Transformer's relevance showed comparable performance across all transformers, with the text-targeted transformer yielding slightly better results (84.6% accuracy, 79.7% macro F1). This indicates effective multimodal information fusion by the Crossmodal Transformer layer, with each modality contributing to the overall performance.

Furthermore, it's noteworthy that the model's performance decreased when trained without the in-game data, dropping to 85.1% accuracy and 80.1% macro F1 score compared to our full model's 87.3% accuracy and 83.0% macro F1 score. This finding underscores the significant role of contextual information (in this case, in-game data) in enhancing the model's performance for turn-taking predictions.

Table 5. Ablation Study Result

| Metric | Acc | Macro F1 | Precision | Recall |
|---|---|---|---|---|
| Text only | 79.9 | 72.7 | 72.3 | 73.3 |
| Vision only | 73.4 | 64.1 | 63.8 | 64.6 |
| Audio only | 79.4 | 72.1 | 71.6 | 72.7 |
| In-game only | 73.3 | 65.7 | 64.9 | 67.2 |
| Only$[V,A,G \to T]$ | 84.6 | 79.7 | 78.5 | 81.4 |
| Only$[T,A,G \to V]$ | 83.5 | 78.0 | 77.1 | 79.4 |
| Only$[T,V,G \to A]$ | 84.3 | 79.4 | 78.1 | 81.1 |
| Only$[T,V,A \to G]$ | 83.6 | 78.9 | 77.4 | 81.3 |
| Without In-game | 85.1 | 80.1 | 79 | 81.4 |

*4.1.3. Model Optimization Result*

To optimize the model for real-time applications, we converted it to TensorFlow Lite (TFLite) format. This conversion resulted in a significant reduction in both model size and inference time while maintaining performance (Table 6). These measurements were specifically taken in an environment equipped with an Intel Core i7-10700 CPU and 32.0GB RAM.

The optimized model's inference time (25.8 ms) allows for rapid response, which is crucial given that typical gaps between turns in human conversations average around 200 ms [29]. This quick inference time helps bridge the gap between AI response capabilities and the swift timing of human conversational turns, even when accounting for the longer latencies involved in language production.

Table 6. Model Optimization Result

|  | Tensorflow model | TFlite model |
|---|---|---|
| Model size [MB] | 6.44 | 2.28 |
| Inference time [ms] | 102.6 | 25.8 |
| Accuracy | 87.3 | 87.3 |
| Macro F1 | 83.0 | 82.9 |

*4.2. User Study Analysis Result*

*4.2.1. Utterance counts*

In our user study experiments with the turn-taking model deployed in real-time, we conducted a detailed analysis of speech dynamics, focusing on the frequency of utterances by both avatars and human participants. Table 7 presents the differences in utterance counts between human and avatar participants across different experimental conditions: the Control condition (model turned off) and the Turn-taking condition (model turned on).

The results in Table 7 reveal that the frequency of utterances remained consistent across both conditions. This observation indicates that the implementation of our turn-taking model did not significantly alter the frequency of speech by either avatars or human participants. The significance of these findings is denoted with symbols (* for $p < 0.05$, ** for $p < 0.01$, *** for $p < 0.001$). In the Korean language experiments, there were no notable differences in utterance frequency between the Turn-taking and Control conditions. A similar pattern was observed with English speakers, with only a minor variation in the frequency of avatar's utterances. In summary, this analysis demonstrates that the integration of the turn-taking model did not substantially impact the overall volume of conversation. This finding underscores the model's capability to facilitate natural turn-taking without disrupting the usual flow of conversation. It also implies that our model seamlessly integrates into human-agent dialogues.

Table 7. Utterance Counts, By Condition

|  | Turn-taking (std) | Control (std) | *p*-val | *Sign.* |
|---|---|---|---|---|
| Korean |  |  |  |  |
| Avatar | 81.13(15.09) | 82.50(20.33) | 0.8216 |  |
| Human | 128.40(53.18) | 105.20(65.19) | 0.2658 |  |
| English |  |  |  |  |
| Avatar | 87.10(18.25) | 103.91(19.69) | 0.0575 |  |
| Human | 151.00(75.32) | 136.91(62.10) | 0.6440 |  |

*4.2.2. Interruption Frequency*

We next examined the frequency of interruptions during conversations in the user study, analyzing this aspect across different languages and experimental conditions for both human participants and the avatar. An interruption was defined as an instance where one speaker talked over another, regardless of whether it was intentional or accidental. These findings are detailed in Table 8. To ensure a fair comparison, we calculated the frequency of interruptions as a percentage of the other speaker's total utterances, since not all individuals talk as much as others. This approach helps in understanding how interruptions by one party affect the speaking opportunities of the other within a fixed time frame (in this case, 30 minutes). A significant variation in this interruption percentage between the experimental conditions could indicate the influence of our turn-taking model on the flow of conversation. Additionally, this metric assists in assessing the avatar's turn-taking proficiency across different linguistic contexts. Ideally, an effective turn-taking model should minimize inappropriate interruptions while maintaining a balanced conversation flow, irrespective of the language used.

From the results presented in Table 8 and Figure 6, several observations were made: First, in both Korean and English experiments under the Turn-taking condition, there was a noticeable reduction in the avatar's interruption ratio compared to the Control condition. This suggests that our model had some impact on turn-taking behavior by the avatar, although the change was not statistically significant. Secondly, on the human side, the interruption ratio for English speakers remained relatively consistent across both conditions, whereas for Korean speakers the interruption ratio in the Turn-taking condition (5.72%) was notably higher than in the Control condition (2.53%).

The disparity in interruption ratios between Korean and English speakers, particularly under the Turn-taking condition, suggests that cultural or linguistic factors may influence the effectiveness of our turn-taking model, e.g. different turn-taking cues across languages. That may suggest we need separately trained models for different languages (see Discussion). Another possible explanation for this disparity though is that the model training data for Korean and English exhibit different ratios of class imbalance (see Section 3.3.5), potentially leading to distinct turn-taking dynamics for each language being obfuscated in the training data. Given the varied results observed between Korean and English speakers, further research into other languages or bilingual speaker settings might enhance our understanding of the utility of turn-taking models cross-linguistically. In summary, these findings suggest that language-specific turn-taking models might be more effective than a universal model for all languages, but more work is needed.

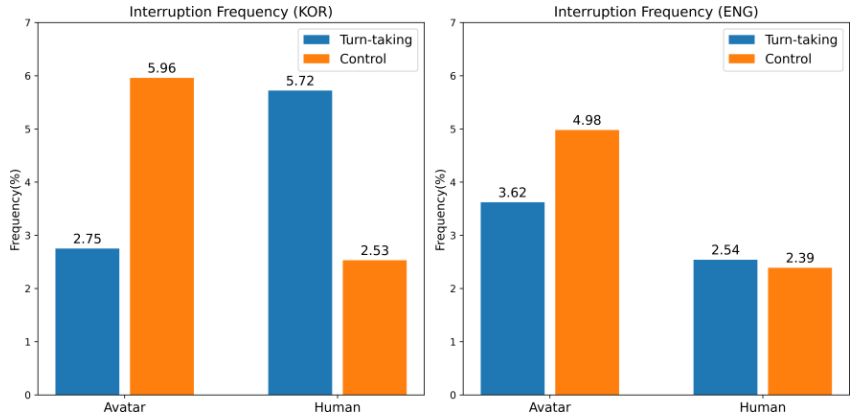

Figure 6: Comparative Visualization of Interruption Frequencies, By Languages and Conditions

Table 8. Interruption Frequency, By Condition

|  | Turn-taking (std) | Control (std) | *p*-val | *Sign.* |
|---|---|---|---|---|
| Korean |  |  |  |  |
| Avatar | 2.75% (0.01) | 5.96% (0.04) | 0.0596 |  |
| Human | 5.72% (0.04) | 2.53% (0.02) | 0.0253 | * |
| English |  |  |  |  |
| Avatar | 3.62% (0.02) | 4.98% (0.03) | 0.2650 |  |
| Human | 2.54% (0.03) | 2.39% (0.03) | 0.8994 |  |

*4.2.3. User perceptions based on Instrument Data*

During the study, we also explored the impact of an artificial agent's turn-taking ability on human perceptions. To do so, we analyzed data collected from HRI instruments, as detailed in Section 3.4.4. The primary instruments for this analysis were the Godspeed Instrument and the Networked Minds Instrument, both designed to measure various dimensions of human perception during interactions with artificial avatars.

Despite incorporating a turn-taking model designed to reduce interruption frequency, our analysis revealed no significant differences in either the Godspeed or Networked Minds measurements between the Turn-taking and Control conditions, as shown in Tables 9 and 10. This observation was consistent across both Korean and English-speaking participants, suggesting that merely reducing interruptions may not be sufficient to enhance user perception and experience with the virtual artificial agent. Another possible explanation for these results is that turn-taking is just one component of natural human communication, and its impact on user perceptions in the context of HAI and HRI may involve additional factors beyond minimizing interruptions. In fact, some level of interruption might contribute to a more natural interaction dynamic, as suggested by Khouzaimi and Laroche [15]. Moreover, the turn-taking ability of the artificial agent, as measured by the instruments used in this study, may not be sufficient to significantly influence human perceptions. It is also important to consider the limitations of the instruments utilized, as they may not be sensitive enough to capture subtle nuances in human perceptions that could be influenced by the turn-taking ability of an artificial agent.

Table 9: Instrument Analysis, By Condition (KOR)

|  | Turn-taking | Control | $p$-val | Sign. |
|---|---|---|---|---|
| Godspeed (total) | 3.15 | 3.26 | 0.1397 |  |
| Subscales |  |  |  |  |
|   Anthropomorphism | 2.63 | 2.68 | 0.7468 |  |
|   Animacy | 3.21 | 3.33 | 0.4312 |  |
|   Likeability | 3.45 | 3.59 | 0.3192 |  |
|   Perc. Intelligence | 3.27 | 3.46 | 0.2352 |  |
|   Perc. Safety | 3.18 | 3.20 | 0.9285 |  |
| NM Self | 3.08 | 3.21 | 0.2337 |  |
| NM Other | 3.15 | 3.17 | 0.8538 |  |

Table 10: Instrument Analysis, By Condition (ENG)

|  | Turn-taking | Control | $p$-val | Sign. |
|---|---|---|---|---|
| Godspeed (total) | 3.39 | 3.45 | 0.5140 |  |
| Subscales |  |  |  |  |
|   Anthropomorphism | 3.10 | 2.98 | 0.5691 |  |
|   Animacy | 2.90 | 3.26 | 0.0454 | * |
|   Likeability | 4.00 | 3.95 | 0.7257 |  |
|   Perc. Intelligence | 3.54 | 3.55 | 0.9760 |  |
|   Perc. Safety | 3.60 | 3.64 | 0.8818 |  |
| NM Self | 3.24 | 3.24 | 0.9631 |  |
| NM Other | 3.21 | 3.32 | 0.4075 |  |

# 5. Discussion

## 5.1. Summary of Results

Our primary objective was to investigate whether there are differences in how people interact with a virtual avatar capable of turn-taking prediction in cooperative game environments, encompassing both model development (benchmarking) and user studies of the deployed model in real-time. In such cooperative environments, effective communication and collaboration between humans and the avatar are essential to achieve a common goal. The key findings from our comprehensive analysis can be summarized as follows.

Our proposed multimodal turn-taking prediction model outperformed baseline models (Early Fusion LSTM, Late Fusion LSTM, and Mult) across all metrics, exhibiting faster convergence during training. Specifically, our model achieved an accuracy of 87.3% and a macro F1 score of 83.0%, surpassing the next best model (Mult) which achieved 82.8% accuracy and 77.0% macro F1 score. The ablation study revealed that text and audio modalities had the most significant impact on turn-taking prediction, while the Crossmodal Transformer layer effectively fused multimodal information. Notably, the inclusion of in-game contextual data enhanced the model's performance, underscoring the importance of context in turn-taking predictions. Optimization efforts, involving conversion to TFLite format, significantly reduced model size (from 6.44MB to 2.28MB) and inference time (from 102.6ms to 25.8ms) while maintaining performance. This optimization brings the model's response time closer to the pace of human conversations, facilitating more natural human-agent dialogues.

In the user study of the deployed model, our analysis revealed several important insights. First, the integration of the turn-taking model into the avatar's speech system demonstrated its effectiveness in maintaining a consistent level of dialogue in the Turn-taking condition, as compared to the Control condition (utterance counts). This consistency indicates that the avatar's turn-taking ability did not disrupt the natural flow of conversation or alter the frequency of dialogue. Second, the turn-taking model slightly reduced the avatar's interruption frequency, although the change was not statistically significant. However, the effects on human speaker interruptions were variable. Korean-speaking participants showed a higher rate of interruptions in the Turn-taking condition compared to the Control condition, but this was not true for English-speaking participants. This suggests that turn-taking dynamics might vary across languages, highlighting the potential need for language-specific turn-taking models, as a universal model may not adequately capture the unique conversational styles and turn-taking cues of different languages. Lastly, the turn-taking ability of the avatar did not significantly influence participants' perceptions of the artificial agent, at least not in a manner discernible by the Godspeed and Networked Minds instruments.

The reasons behind these findings are not entirely clear but may be attributed to cultural differences in social norms governing appropriate interaction behavior between East Asian and Western cultures, as well as possible differences in turn-taking cues during speech interactions [28, 29, 30]. Further research is needed to explore how such turn-taking models can be incorporated into conversational user interfaces (CUI) going forward, particularly those that are intended to operate across multiple languages (or even dialects of a single language) [58].

*5.2. Implications*

Turn-taking models, such as the one we proposed in this research, hold tremendous promise for predicting enhancing HAI and CUI via more fluid speech interactions. Our approach in particular stands out due to its incorporation of contextual data about the situation, a factor often overlooked in traditional turn-taking studies that typically focus only on text, audio, and visual data [59, 60]. In our case, such contextual data related to the game environment, but that could easily be extended to other real-world interaction environments. This inclusion of context allows for a deeper understanding of interaction dynamics in complex, task-oriented environments, which enhances the utility of both verbal and non-verbal cues.

There are also some design implications related to our choice of model architecture, which diverges from other previous models in critical ways that merit further discussion (and research) by the scientific community. One key aspect of our methodology is the utilization of the Crossmodal Transformer architecture. That approach enables the effective integration of multiple types of data within a single model that can be trained using widely available hardware resources [47]. Doing so provides a more comprehensive understanding of the interaction context, enhancing the utility verbal and non-verbal cues by considering them simultaneously rather than in isolation, whereas the interpretation of a verbal cue often depends on non-verbal cues (and vice versa). In other words, we may be able to improve turn-taking behavior, by integrating information of those cues *within* the model architecture explicitly in order to make models that are applicable across a variety of scenarios without the need for different models for each individual scenario [61, 62].

We also note that the efficiency of our modeling approach renders it particularly suitable for real-time applications, broadening its potential impact and practical use in diverse environments, such as homes and workplaces [63, 64]. For instance, in a healthcare setting, a robot equipped with our turn-taking model could engage more naturally with patients, facilitating the collection of health information and providing comfort, thereby potentially improving patient care [65].

There is also potential here to use turn-taking models to augment the capabilities of existing speech systems and large language models (LLMs), which have some weaknesses noted in recent literature when used for conversational systems (e.g. staying on-topic, appropriate turn-taking) [66,67]. For instance, our approach is designed to complement existing ASR systems, such as Microsoft Azure. While traditional ASR systems primarily rely on detecting speech pauses to determine turn endings [12, 13, 14], our method augments this by continuously monitoring the speaker's speech patterns to predict opportunities for the agent to speak ahead-of-time. This addresses some of the limitations associated with pause-based speech recognition, e.g. the need for manual post-cleanup of automatic annotations which is impractical for deployed systems. Additional concerns related to the lack of context-awareness by LLMs [68] may also be partially addressed through incorporation of context into models for turn-taking, among others.

*5.3. Limitations*

In this section, we address the limitations of our study and suggest potential directions for future research. Our turn-taking model relies on a predefined one-second threshold to predict upcoming turn-taking events, based on data from the preceding five seconds. This threshold was strategically selected to balance the need for real-time data processing with the model's predictive accuracy. However, a fixed one-second threshold might not encompass the full spectrum of turn-taking dynamics, especially in scenarios involving rapid or subtle conversational shifts. Future iterations of the model could benefit from exploring shorter threshold settings to enhance detection and prediction of quicker turn-taking instances.

Another limitation of our proposed approach is its inability to effectively recognize backchannel events, which are often brief and can be non-verbal or verbal acknowledgments occurring within short time frames. Indeed, backchanneling is known to be an important feature of human-human dialogue. Future research could aim to improve our proposed model's capability to detect and interpret both turn-taking and backchannel events.

For assessing participant perceptions of the virtual avatar and interaction quality, we utilized the Godspeed and Networked Minds questionnaires. While these instruments provide valuable insights, a survey instrument specifically designed for assessing turn-taking dynamics would likely yield more detailed findings. Such an instrument could offer a deeper understanding of the nuances in HRI, particularly in the context of our study.

Lastly, our approach is primarily tailored for dyadic interactions and has not been tested in multi-party conversation contexts. Multi-party interactions introduce distinct dynamics, as discussed in existing literature [8, 16, 17, 22, 23]. The complexity of turn-taking escalates in multi-party settings due to the involvement of multiple speakers and listeners, each potentially assuming different roles in the conversation. In such environments, the coordination of speaking turns relies not only on verbal cues but also more significantly by non-verbal signals like gaze direction, facial expressions, and body language.

Therefore, while our model demonstrates promise in one-on-one scenarios, its applicability and effectiveness in group conversations remain to be explored.

*5.4. Future Work*

In our future research, we plan to explore several promising directions. These include refining our turn-taking model by optimizing or varying the threshold for quicker turn-taking instances, incorporating backchannel recognition, and developing specialized survey instruments dedicated to assessing turn-taking dynamics. Additionally, applying the model in real-world environments, particularly in multi-party conversations, will be crucial to evaluate its practical efficacy. We also plan to integrate certain significant non-verbal cues, such as eye gaze and gestures, into our turn-taking prediction model in a more explicit manner. These non-verbal cues, including gazing away, inhaling, or initiating a gesture, often signal a human speaker's intent to initiate or yield a turn [24]. Emphasizing those specific cues could potentially enhance the model's ability to predict turn-taking events, enriching conversational dynamics.

Another intriguing avenue is to extend our model to additional languages, such as Japanese or German, as well as to bilingual speakers. While general turn-taking principles are somewhat universal across languages [30], there are notable differences in specific cues used and overall distributions. Testing our model in these varied linguistic environments could provide valuable insights into its universality and adaptability. Moreover, we could potentially explore applications of this technology to humans, such as $2^{nd}$ language learning in children or older adults with communication difficulties due to cognitive decline (e.g. dementia).

Lastly, deploying our turn-taking model in physical robots presents an exciting research frontier. In multi-party interactions, physical robots may offer more efficient turn-taking management compared to virtual agents on 2D displays [22]. The physical presence of a robot, even in dyadic interactions, has been shown to be beneficial in contexts such as language learning, compared to virtual characters on a screen [23]. This suggests that physical robots could provide superior opportunities for social interaction compared to virtual agents or voice assistants. Undertaking these initiatives aims not only to address the current limitations of our approach but also to pioneer advancements in the nuanced realms of HAI and HRI. By pushing the boundaries of turn-taking capabilities, we aim to facilitate more natural, intuitive, and effective communication between humans and artificial agents. This endeavor will pave the way for innovative applications that enhance everyday human-technology interactions.

## 6. Acknowledgments

This work was primarily supported by a grant from the National Research Foundation of Korea (NRF) (Grant number: 2021R1G1A1003801). Additional support was provided by Institute of Information & communications Technology Planning & Evaluation (IITP) grant funded by the Korea government (MSIT) (RS-2022-00143911, AI Excellence Global Innovative Leader Education Program). We would like to thank our other research collaborators who contributed to this work.